\documentclass[aps,prd,twocolumn,showpacs,groupedaddress,floatfix,letter]{revtex4-1}

\usepackage{graphicx}
\usepackage[symbol*]{footmisc}
\usepackage{array}
\usepackage{amsmath}
\usepackage{amssymb}
\usepackage{amsfonts}

\DeclareGraphicsExtensions{.eps, .eps, .jpg, .png, .pdf}

\def\cF{\mathcal{F}}
\def\cA{\mathcal{A}}

\newcommand{\de}{\mathrm{d}}
\newcommand{\BesselI}[2]{\mathrm{I}_#1(#2)}
\newcommand{\BesselK}[2]{\mathrm{K}_#1(#2)}

\begin{document}

\title{Dynamical Four-Form Fields}

\author{Ufuk Aydemir}
\email[]{uaydemir@physics.umass.edu}
\author{Luca Grisa}
\email[]{lgrisa@physics.umass.edu}
\author{Lorenzo Sorbo}
\email[]{sorbo@physics.umass.edu}

\affiliation{Department of Physics, UMass Amherst}

\date{\today}

\begin{abstract}
We present an example of a gauge-invariant Lagrangian that contains four derivatives and describes one massive, non-ghostlike, degree of freedom.
\end{abstract}


\maketitle

\paragraph*{Motivation.} Gauge invariance is often assumed to imply the existence of massless degrees of freedom, although there are famous counter-examples, ranging from~\cite{Schwinger:1962tn,Schwinger:1962tp} to~\cite{Deser:1982vy,Porrati:2001gx}. Moreover, it is often assumed that theories with higher derivatives contain ghosts or other pathologies. Recently,~\cite{Deser:2009hb} has observed that the model, proposed in~\cite{Bergshoeff:2009hq} and therein called ``new massive gravity,'' does provide a counter-example to both these commonly accepted no-goes. Alas, the model~\cite{Bergshoeff:2009hq} is defined only in three-dimensional space-times.

In the present paper we will discuss a model that violates both aforementioned assumptions in any number of dimensions. As we will see, the model shares many of the properties with the new massive gravity of~\cite{Bergshoeff:2009hq}, but without the limitation of three space-time dimensions (though we will mostly focus, for the sake of argument, on four dimensions).

The key ingredient of this model is, in $d$ space-time dimensions, a ($d-1$)-form field potential $\cA_{\mu_1\mu_2\cdots\mu_{d-1}}$. As for electromagnetism, which is described by a 1-form potential, we require gauge invariance under the transformation $\cA_{\mu_1\mu_2\cdots\mu_{d-1}}\rightarrow \cA_{\mu_1\mu_2\cdots\mu_{d-1}}+\partial_{[\mu_1}{\cal {B}}_{\mu_2\cdots\mu_{d-1}]}$, where ${\cal {B}}_{\mu_1\cdots\mu_{d-2}}$ is an arbitrary ($d-2$)-form. Thus, the field strength $\cF_{\nu\mu_1\mu_2\cdots\mu_{d-1}}\equiv \partial_{[\nu}\cA_{\mu_1\mu_2\cdots\mu_{d-1}]}$ is gauge invariant.

Specializing to four space-time dimensions, it is usually assumed, by analogy with electromagnetism, that the Lagrangian for a 4-form field strength should be
\begin{equation}\label{lagr0}
\mathcal{L}=\sqrt{-g}\Big[-\frac{1}{48}\,\cF^{\mu\nu\rho\lambda}\,\cF_{\mu\nu\rho\lambda}\Big]\,.
\end{equation}
Had this Lagrangian -- or, more generally, any polynomial in $\cF_{\mu\nu\rho\lambda}$ --  been chosen, the field $\cA_{\mu\nu\rho}$ would not have any local dynamics. We observe that the tensorial structure of the field strength imposes $\cF_{\mu\nu\rho\lambda}(x^\alpha)=\sqrt{-g}\,\epsilon_{\mu\nu\rho\lambda}\,\Phi(x^\alpha)$, where $\Phi(x^\alpha)$ is a scalar field and $\epsilon_{\mu\nu\rho\lambda}$ the Levi-Civita symbol with $\epsilon_{0123}=1$. The equations of motion derived from \eqref{lagr0}, $\nabla^\mu\,\cF_{\mu\nu\rho\lambda}=0$, imply $\partial_\mu\Phi=0$, {\it i.e.}, $\Phi$ is constant.

We will show that adding higher derivatives term, $\nabla^\alpha \cF_{\mu\nu\rho\lambda}$, in the Lagrangian allows the (dual) scalar field $\Phi$ to propagate. Moreover, for an appropriate choice of parameters, the propagating degree of freedom has a well-behaved kinetic term and a positive-definite potential.

\paragraph*{The Lagrangian.} The most general Lagrangian, quadratic in $\cF$ and with at most four derivatives and ``dimension six'' operators, is (we will assume the ``mostly plus" convention for the signature of the metric)
\begin{align}
\label{laglong}
\mathcal{L}=&\sqrt{-g}\Big[
	\frac{c_0}{48}\cF^{\mu\nu\rho\lambda}\cF_{\mu\nu\rho\lambda}+
	\frac{c_1}{48\,M^2}\nabla^\alpha\cF^{\mu\nu\rho\lambda}\nabla_\alpha\cF_{\mu\nu\rho\lambda}+\nonumber\\
	+&\frac{c_2}{12\,M^2}\nabla_\alpha\cF^{\alpha\nu\rho\lambda}\nabla^\beta\cF_{\beta\nu\rho\lambda}+
	{\mathrm {terms\ in\ }}R\cF\cF\Big]\,,
\end{align}
where $R\cF\cF$ indicates any term built by contracting two field strengths with a Riemann tensor or its contractions. We will neglect these terms to ease the discussion, but their role will be discussed at the end of this note. In~\eqref{laglong}, the coefficients $c_0,\,c_1,\,c_2$ are dimensionless numbers, and $M$ is a mass scale. A term proportional to $\nabla^\alpha \cF^{\beta\nu\rho\lambda}\nabla_\beta \cF_{\alpha\nu\rho\lambda}$ can also be added to~\eqref{laglong}, however it is identical to the term proportional to $c_{2}$ after an integration by parts and so it will be neglected.

It is convenient to rescale $\cA_{\mu\nu\rho}$ by a factor of $M$; from now on, we will work with the field $A_{\mu\nu\rho}\equiv\cA_{\mu\nu\rho}/M$ (and $F_{\mu\nu\rho\lambda}\equiv\cF_{\mu\nu\rho\lambda}/M$). The equations of motion derived by varying the Lagrangian~\eqref{laglong} with respect to $A_{\nu\rho\lambda}$ can be rearranged to take the form
\begin{equation}
\nabla^\mu\left[c_0\,M^2\,F_{\mu\nu\rho\lambda}-\left(c_1+c_2\right)\,\nabla^\alpha\,\nabla_\alpha\,F_{\mu\nu\rho\lambda}\right]=0\,,
\end{equation}
where we have used the fact that $F_{\mu\nu\rho\lambda}$ is totally antisymmetric in its four indices. Setting $F_{\mu\nu\rho\lambda}=\sqrt{-g}\,\epsilon_{\mu\nu\rho\lambda}\,\phi$, the equation of motion becomes
\begin{equation}\label{eomphi}
\left(c_1+c_2\right)\,\nabla^\alpha\,\nabla_\alpha\,\phi-c_0\,M^2\,\phi=q\,,
\end{equation}
where $q$ is an integration constant. Therefore, the Lagrangian~\eqref{laglong} describes one dynamical scalar degree of freedom of mass
\begin{equation}
\mu^2\equiv\frac{c_0}{c_1+c_2}\,M^2\,,
\label{mass}
\end{equation}
and vacuum expectation value proportional to the integration constant $q$.
In other words, our system is equivalent to a landscape of massive scalars characterized by different values of the integration constant $q$.

The stress-energy tensor obtained by varying the Lagrangian~\eqref{laglong} with respect to $g_{\mu\nu}$ reads, in terms of the scalar $\phi$, 
\begin{align}
T_{\mu\nu}=\,
	&(c_1+c_2)\left[
		\nabla_\mu\phi\,\nabla_\nu\phi-g_{\mu\nu}\left(\frac{1}{2}\nabla^\alpha\phi\nabla_\alpha\phi+
		\phi\,\Box\phi\right)\right]+\nonumber\\
	+&g_{\mu\nu}\,\frac{c_0}{2}\,M^2\,\phi^2\,,
\end{align}
or, by using \eqref{eomphi}, as
\begin{align}
T_{\mu\nu}=&
	(c_1+c_2)\,\nabla_\mu\phi\nabla_\nu\phi+\\
	-&g_{\mu\nu}\left(\frac{c_1+c_2}{2}\nabla^\alpha\phi\nabla_\alpha\phi+\frac{c_0}{2}\,M^2\,\phi^2+q\,\phi\right)\,.\nonumber
\end{align}

The requirement that the stress-energy tensor is positive definite imposes the conditions $c_1+c_2>0$ (equivalent to the no-ghost condition for the scalar kinetic term) and $c_0>0$ (equivalent for the mass of $\phi$ to be non-tachyonic).

We note that the requirement of non-tachyonic mass term corresponds to the ``wrong'' sign for the $F^2$ term. This is precisely the same phenomenon observed in the new massive gravity model of~\cite{Bergshoeff:2009hq}, where the role of $F^2$ is played by the Ricci scalar. Let us also note, however, that the addition in the Lagrangian of higher powers of the form field, {\it e.g.}, $F^{\alpha\beta\gamma\delta}F^{\mu\nu\rho\lambda}F_{\alpha\beta\rho\lambda}F_{\mu\nu\gamma\delta}$, can stabilize the vacuum of the theory at non-vanishing values of the form field even if $c_0<0$.

The properties of our system become more apparent in the first order formalism, where the potential $A_{\mu\nu\rho}$ and the field strength $F_{\mu\nu\rho\lambda}$ are treated as independent variables, and the condition $F_{\mu\nu\rho\lambda}\equiv \partial_{[\mu}A_{\nu\rho\lambda]}$ is imposed by introducing a Lagrange multiplier $q(x^\alpha)$. The Lagrangian then reads
\begin{align}
\label{laglagr}
\mathcal{L}=\sqrt{-g}\Big[&
	\frac{c_0\,M^2}{48}\,F^{\mu\nu\rho\lambda}\,F_{\mu\nu\rho\lambda}+
	\frac{c_1}{48}\nabla^\alpha F^{\mu\nu\rho\lambda}\,\nabla_\alpha F_{\mu\nu\rho\lambda}+\nonumber\\
	&+\left.\frac{c_2}{12}\nabla_\alpha F^{\alpha\nu\rho\lambda}\,\nabla^\beta F_{\beta\nu\rho\lambda}\right]+\nonumber\\
	&+\frac{q}{24}\,\epsilon^{\mu\nu\rho\lambda}\,\left(F_{\mu\nu\rho\lambda}-4\,\partial_{\mu}A_{\nu\rho\lambda}\right)\,.
\end{align}

Since $F^{\mu\nu\rho\lambda}$ is now a fundamental field, we can replace it by $\sqrt{-g}\,\epsilon_{\mu\nu\rho\lambda}\,\phi(x^\alpha)$ directly in the Lagrangian:
\begin{align}\label{laglagrphi}
\mathcal{L}=&\sqrt{-g}\Big[-\frac{c_1+c_2}{2}\nabla^\alpha \phi\,\nabla_\alpha \phi-\frac{c_0}{2}\,M^2\,\phi^2\,-q\,\phi\Big]+\nonumber\\
-&\frac{q}{6}\,\epsilon^{\mu\nu\rho\lambda}\,\partial_{\mu}A_{\nu\rho\lambda}\,.
\end{align}

The equation for $A^{\mu\nu\rho}$, $\partial_\lambda q=0$, simply implies the Lagrange multiplier $q$ to be a constant.

Despite this system being very similar to the one described in~\cite{Kaloper:2008qs,Kaloper:2008fb}, it is not the same: while in~\cite{Kaloper:2008qs,Kaloper:2008fb} the different vacua have all the same cosmological constant, in our case they do not, since the vacuum energy is proportional to $q^{2}$. Further differences emerge when we couple our form field to matter.

\paragraph*{Coupled to matter.} Given the ``wrong'' sign of the term $F^2$, an interesting question to answer is whether this model is stable once coupled to matter. $p$-form fields couple naturally to ($p-1$)-branes. In four dimensions, a 2-brane sources a 3-form field
\begin{align}
\mathcal{L}_{\mathrm {brane}}=\int_{\partial}\de^3\xi\,
         \Big[-\sigma\sqrt{-\hat g}
         	+\frac{e}{6}A_{\mu\nu\rho}
         	\frac{\partial z^{\mu}}{\partial\xi^{a}}
			\frac{\partial z^{\nu}}{\partial\xi^{b}}
			\frac{\partial z^{\rho}}{\partial\xi^{c}}\epsilon^{abc}\Big]\,,
\label{lagrcoupling}
\end{align}
where $e$ is the charge and $\sigma$ the tension of the membrane; we denote by $\partial$ the brane world-volume. Note that $e$ is proportional by a factor $M$ to the original brane charge for the field $\cA_{\mu\nu\rho}$. This coupling could destabilize the vacuum through brane nucleations, described by the following equations of motion (neglecting the role of gravity)
\begin{align}
	&z^\nu_{,a}\,\partial_\nu(z^\mu_{,a})=
		\frac{e}{6\,\sigma}\,\phi\,\epsilon^{\mu\nu\rho\lambda}\,
		\frac{\partial z_{\nu}}{\partial\xi^{a}}
		\frac{\partial z_{\rho}}{\partial\xi^{b}}
		\frac{\partial z_{\lambda}}{\partial\xi^{c}}\,\epsilon^{abc}\,,\label{EoMe}\\
	&\epsilon^{\mu\nu\rho\lambda}\,\partial_{\mu}q=
		-e\int\de^3\xi\,\delta^{(4)}\left(x-z(\xi)\right)\,
		\frac{\partial z^{\nu}}{\partial\xi^{a}}
		\frac{\partial z^{\rho}}{\partial\xi^{b}}
		\frac{\partial z^{\lambda}}{\partial\xi^{c}}\epsilon^{abc}\,,\label{EoMA}\\
	&\Box\,\phi-\mu^2\phi-\frac{q}{c_1+c_2}=0\,,
	\label{EoMphi}
\end{align}
where we have already substituted the 4-form curvature of $A_{\mu\nu\rho}$ with $F_{\mu\nu\rho\lambda}=\epsilon_{\mu\nu\rho\lambda}\,\phi$ and $\mu^{2}$ is the effective mass found in eq.~\eqref{mass}.

The solution for the equation of motion for the 3-brane \eqref{EoMe} is that of a hyperboloidal bubble of constant ``radius'' $r_0$. From \eqref{EoMA}, we notice that the Lagrange multiplier $q$ is constant away from the bubble and changes by $e$, the charge of the 2-brane, when crossing the bubble, {\it i.e.}, $|\Delta q|=e$.

Since we are interested in instanton solutions for the nucleation of bubbles with different value of $q$, we will consider the Euclidean action $\mathcal{S}_\mathrm{E}$ of \eqref{laglagrphi} with coupling \eqref{lagrcoupling} on the solution of the equations of motion \eqref{EoMe}, \eqref{EoMA} and \eqref{EoMphi}
\begin{align}
	\mathcal{S}_\mathrm{E}=&\int\de^4x\left[
		\frac{1}{2}(\partial\bar\phi)^2+\frac{1}{2}\mu^2\bar\phi^2+\frac{q}{\tilde{c}}\bar\phi\right]+\nonumber\\
		&+\sigma\int_{\partial}\de^3\xi\sqrt{\hat g}\,,
	\label{actionE}
\end{align}
where $\bar\phi$ is the canonically normalized scalar field $\bar\phi\equiv\tilde{c}\,\phi$ with $\tilde{c}\equiv \sqrt{c_{1}+c_{2}}$, and where we have taken into account the appropriate boundary terms in the Lagrangian. The function $q(r)$ is given by $q(r)= q_{<}\,\theta(r_0-r)+q_{>}\,\theta(r-r_0)$, where the step function $\theta(r)$ is equal to 1 when its argument is positive and equal to zero when it is not, and $|q_{>}-q_{<}|=e$. By using the spherical symmetry of the problem, we can simplify \eqref{actionE} even further
\begin{align}
	\mathcal{S}_\mathrm{E}=
		\pi^2\int\de r\,r^3\left[
		(\partial_r\bar\phi)^2+\mu^2\bar\phi^2+2\,\frac{q(r)}{\tilde{c}}\bar\phi\right]+2\,\pi^2\sigma\,r_0^3
	\label{actionEint}
\end{align}
and the equation of motion for $\bar\phi$ becomes
\begin{equation}
    \bar\phi''+3\frac{\bar\phi'}{r}-\mu^2\bar\phi-\frac{q(r)}{\tilde{c}}=0\,.
\end{equation}
Regularity for both $r\rightarrow0^+$ and $r\rightarrow\infty$ constrains the solution to be
\begin{align}
\bar\phi(r)=\left\{
\begin{array}{ll}
	-\frac{q_{<}}{\mu^2\,\tilde{c}}+A\,\frac{\BesselI{1}{\mu\,r}}{\mu\,r} & {\mathrm {,\ \ for}}\ r<r_0\\
	-\frac{q_{>}}{\mu^2\,\tilde{c}}+B\,\frac{\BesselK{1}{\mu\,r}}{\mu\,r} & {\mathrm {,\ \ for}}\ r>r_0
\end{array}
\right.
\end{align}
where $\BesselI{n}{\mu\,r}$ and $\BesselK{n}{\mu\,r}$ are Bessel functions. The constants $A$ and $B$ are found by requiring continuity of the function and its derivative across the bubble at $r=r_0$:
\begin{align}
    &A = \frac{q_<-q_>}{\tilde{c}}\,r_{0}^{2}\,\BesselK{2}{\mu\,r_{0}}\\
    &B = -\frac{q_<-q_>}{\tilde{c}}\,r_{0}^{2}\,\BesselI{2}{\mu\,r_{0}}\,,
\end{align}
The difference between the action for the solution of constant $q_{>}$ and the action for $q(r)$ is
\begin{align}
    \Delta \mathcal{S}_\mathrm{E}&=2\,\pi^2\,\sigma\,r_0^3+\frac{\pi^2\,r_0^4}{4\,\mu^2}\,\frac{q_>^2-q_<^2}{\tilde{c}^2}+\nonumber\\
&+\frac{\pi^2\, r_0^4}{\mu^2}\,\frac{\left(q_<-q_>\right)^2}{\tilde{c}^2}\,\BesselI{2}{\mu\, r_0}\,\BesselK{2}{\mu\, r_0}\,.
\label{DeltaSE}
\end{align}

This expression is not very illuminating; it is convenient to consider the limits  $\mu\,r_0\ll 1$ and $\mu\,r_0\gg 1$.

In the first limit, the difference in actions reduces to
\begin{equation}\label{rsmall}
    \Delta \mathcal{S}_\mathrm{E}\sim 2\,\pi^2\,\sigma\,r_0^3+\frac{\pi^2\,r_0^4}{2\,\mu^2}\,\frac{q_>\,\varepsilon\,e}{\tilde{c}^2},
\end{equation}
where we have introduced the symbol $\varepsilon=\pm 1$, $q_>-q_<=\varepsilon\,e$, to discriminate between a brane and an anti-brane nucleation. This expression for the change in action can be easily found by direct calculation, observing that, since the bubble is much smaller than the scalar Compton wavelength, $\bar\phi$ stays essentially constant at the value it assumes outside the bubble, {\it i.e.}, $\bar\phi=-q_>/(\mu^2\tilde{c})$.

It is clear that the action~\eqref{rsmall} can always be minimized, with an appropriate choice of the sign $\varepsilon$, at
\begin{eqnarray}
&&r_0\sim 3\,\frac{\sigma\,\mu^2\,\tilde{c}^2}{\left|q_>\right|\,e}\,,\\
&&\Delta \mathcal{S}_\mathrm{E}(r_0)\sim\frac{27\,\pi^2}{2}\,\frac{\sigma^4\,\mu^6\,\tilde{c}^6}{\left|q_>\right|^3\,e^3}\,\,.
\end{eqnarray} 

In the opposite regime $\mu\,r_0\gg 1$, the action reads
\begin{equation}\label{rlarge}
    \Delta \mathcal{S}_\mathrm{E}\sim2\,\pi^2\,\sigma\,r_0^3+\frac{\pi^2}{4\,\mu^2}\,\frac{2\,q_>\,\varepsilon\,e-e^2}{\tilde{c}^2}r_0^4
    +\frac{\pi^2}{2\,\mu^3}\,\frac{e^2}{\tilde{c}^2}r_0^3\,,
\end{equation}
where the second term corresponds to the change in bulk energy induced by the bubble nucleation, whereas the third term represents a correction to the effective membrane tension induced by the gradient in the field $\phi$.

The bubble radius is found by extremizing the action with respect to $r_{0}$. Inspection of eq.~\eqref{DeltaSE}, and of its explicit limits~\eqref{rsmall} and \eqref{rlarge}, shows that it is always possible to find a $r_0>0$ that minimizes the effective action. This implies that, unlike the usual Brown-Teitelboim model~\cite{Brown:1987dd,Brown:1988kg}, this system does not have any absolutely stable vacuum. The reason for such an instability lies in the ``wrong'' sign for the form kinetic term, $F^{2}$. Nevertheless, it is easy to find parameters for which the decay rate, proportional to $e^{-{\cal {S}}_E}$, is extremely small and the system can be considered stable for all practical purposes.

In the decoupling limit for the field $\phi$, $M\rightarrow \infty$ with $q/M\equiv \bar{q}$ and $e/M\equiv \bar{e}$ fixed, the action describing membrane nucleation reads
\begin{equation}\label{rlargesimp}
    \Delta \mathcal{S}_\mathrm{E}=2\,\pi^2\,\sigma\,r_0^3+\frac{\pi^2\,r_0^4}{4\,c_0}\,\left(2\,\bar{q}_>\,\varepsilon\,\bar{e}-\bar{e}^2\right)\,\,.
\end{equation}
which gives the Brown-Teitelboim result~\cite{Brown:1987dd,Brown:1988kg} when $c_{0}=-1$.

By solving numerically the equation $\partial_{r_{0}}\Delta S_{\mathrm{E}}=0$, we find that the probability to nucleate a bubble of $q_{<}=q_{>}-e$ is always smaller than the one corresponding to the Brown-Teitelboim  limit, $M\to\infty$. This can be understood by considering that, when the field $\bar\phi$ is dynamical and massive, a part of the energy gained by nucleating a bubble goes into exciting $\bar\phi$-modes and the smaller the mass, the more likely these modes are to be excited.

Equivalently, it can be noted that $\bar\phi$ generates an effective attractive force between branes. For two parallel branes with charges $e$ and $e'$ placed at a distance $d$, the field $\bar\phi$ mediates a force per unit surface
\begin{equation}
  \frac{F}{s}=\frac{e\,e'}{2\,\mu\,\tilde{c}^2}\,\exp\left[-\mu\,d\right]\,,
\end{equation}
that is, two branes of same charge repel each other, two of opposite charge attract each other. To defy this force the branes will have to be nucleated at radius $r_{0}$ greater than when such a force is absent, {\it i.e.}, $\mu\rightarrow\infty$, the Brown-Teitelboim case. A larger bubble will be less likely to nucleate than a smaller one.

\paragraph*{Schwinger model with a dynamical two-form.} The two dimensional version of our scenario is especially interesting. In fact, in two dimensions, it is known, \cite{Schwinger:1962tp}, that once a massless fermion -- the analog of a light brane in higher dimensions -- of charge $e$ is coupled to a vector field, a gauge-invariant mass term is generated
\begin{equation}
\delta \mathcal{L}_{\mathrm {mass}}=-\frac{e^2}{2\,\pi}\,\cA_{\mu}\,\left(\eta^{\mu\nu}-\frac{\partial^\mu\,\partial^\mu}{\partial^2}\right)\cA_\nu\,.
\end{equation}

This mechanism works also when higher derivatives operators are considered -- as it can be easily checked -- therefore we simply add this gauge-invariant mass to the two dimensional version of the Lagrangian~\eqref{laglong}. The longitudinal mode  of $\cA_{\mu}$ drops out, since the entire Lagrangian is gauge invariant and it depends only on the transverse mode $\cA^{\mathrm{T}}_{\mu}$
\begin{align}
    \mathcal{L}=\,&\frac{c_{0}}{4}\cF_{\mu\nu}^2+\frac{c_{1}}{4\,M^2}(\nabla_\alpha \cF_{\mu\nu})^2
    +\frac{c_2}{2\,M^2}(\nabla^\alpha \cF_{\alpha\mu})^2+\nonumber\\
    &-
    \frac{e^2}{2\pi}\cA^\mathrm{T}_\mu\,\cA^\mathrm{T}{}^\mu+\frac{\mathcal{Q}}{2}\,\epsilon^{\mu\nu}(\cF_{\mu\nu}-2\,\partial_\mu \cA^\mathrm{T}_\nu)\,,
\end{align}
where we have added a Lagrange multiplier $\mathcal{Q}$ to treat the vector field and the curvature independently; $\cF_{\mu\nu}$ is proportional to the two dimensional Levi-Civita symbol ($\cF_{\mu\nu}\equiv\epsilon_{\mu\nu}\Phi$) and $\cA^\mathrm{T}_\mu$ can be integrated out in favor of $\mathcal{Q}$, $\cA^\mathrm{T}_\mu=-(\pi/e^2)\,\epsilon_{\mu\alpha}\partial^\alpha \mathcal{Q}$. These substitutions lead to the following effective Lagrangian in terms of the scalar fields $\Phi$ and $\mathcal{Q}$,
\begin{align}
    \mathcal{L}^\mathrm{eff}=&-\frac{c_1+c_2}{2\,M^2}(\partial_\alpha\Phi)^2-\frac{c_0}{2}\,\Phi^2-\mathcal{Q}\,\Phi
    -\frac{\pi}{2\,e^2}(\partial_\alpha \mathcal{Q})^2\,;
\end{align}
the mass term for the vector field realized via the Schwinger mechanism effectively generates a kinetic term for the Lagrange multiplier $\mathcal{Q}$, making it dynamical. The limit $e\rightarrow0$ decouples $\mathcal{Q}$ from the other fields; the condition $\partial_\mu \mathcal{Q}=0$ is recovered and $\mathcal{Q}$ ceases to propagate.

Before canonically normalizing the fields, we notice that, for $c_1+c_2>0$, the sign of the kinetic term for both $\Phi$ and $\mathcal{Q}$ is that of a healthy scalar field, that is not a ghost.

Let us substitute $\bar\phi\equiv\sqrt{c_1+c_2}\,\Phi/M$ and $q\equiv\sqrt{\pi}\mathcal{Q}/e$
\begin{align}
	\mathcal{L}^\mathrm{eff}=
	&-\frac{1}{2}(\partial_\mu\bar\phi)^2-\frac{1}{2}(\partial_\mu q)^2-\frac{1}{2}\mu^2\bar\phi^2+\nonumber\\
	&-\frac{e\,\mu}{\sqrt{\pi\,c_{0}}}\, q\,\bar\phi\,,
\end{align}
where $\mu$ is the effective mass defined in \eqref{mass}. The model does therefore contain two propagating degrees of freedom of masses:
\begin{equation}
    \mu_\pm^2=\frac{\mu^{2}}{2}
    \left(1\pm\sqrt{1+\frac{4\,e^2}{\pi c_0\,\mu^{2}}}\right)\,.
\end{equation}
Because the coefficient $c_0$ and the effective mass $\mu^{2}$ are positive definite, the square root is always greater than $1$, hence one scalar mode is always tachyonic. This tachyonic instability is the perturbative counterpart of the non-perturbative instability for brane nucleation described above. 

In the limit $\mu^2\gg e^2/c_0$, the two masses are approximately $\mu^2$ and $-e^2/(\pi\,c_0)$ and the tachyonic instability has a characteristic time much longer than timescale $\mu^{-1}$. In the opposite regime $\mu^2\ll e^2/c_0$, the mass eigenvalues have equal magnitude $\pm 2\,e\,\mu/\sqrt{\pi\,c_0}$, signaling a fast instability.

\paragraph*{Non-minimal coupling to gravity.} Let us then discuss the possibility that the Lagrangian~\eqref{laglong} contains couplings of ${\cal {F}}_{\mu\nu\rho\lambda}$ to the Riemann tensor. The formalism, that led to the equivalent expression~\eqref{laglagrphi}, shows that such extra terms would sum up to a coupling proportional to $\phi^2\,R$, with $R$, the Ricci scalar. Therefore, the effective Lagrangian for the canonically normalized field $\bar\phi$ will read
\begin{equation}\label{nonmin}
\sqrt{-g}\left[\left(\frac{M_P^2}{2}-\frac{\xi}{2}\,\bar\phi^2\right)R-\frac{1}{2}\nabla^\alpha\bar\phi\,\nabla_\alpha\bar\phi-\frac{\mu^2}{2}\,\bar\phi^2-\frac{q}{\tilde{c}}\,\bar\phi\right],
\end{equation}
where the value of the parameter $\xi$ depends on the coefficients of the $R\,{\cal {F}}\,{\cal {F}}$ terms in the Lagrangian~\eqref{laglong}. In~\eqref{nonmin} we have also included the Einstein-Hilbert term for gravity, while we have already set $q$ to be a constant, hence the term proportional to $q\,\partial_{\mu}A_{\nu\rho\lambda}$ is neglected.

Direct couplings of ${\cal {F}}_{\mu\nu\rho\lambda}$ to the Riemann tensor, therefore, lead to a non-minimally coupled scalar field. The behavior of $\bar\phi$ is more transparent if we work in the Einstein frame. In order to do so, we redefine the metric $\tilde{g}_{\mu\nu}=\Omega^2\,g_{\mu\nu}$, $\Omega^2\equiv 1-\xi\,\bar\phi^2/M_P^2$. The transformed Lagrangian is 
\begin{equation}
\sqrt{-\tilde{g}}\left[\frac{M_P^2}{2}\,\tilde{R}-\frac{\Omega^2+6\,\xi^2\,\bar{\phi}^2/M_P^2}{2\,\Omega^4}\tilde{\nabla}^\alpha\bar\phi\,\tilde{\nabla}_\alpha\bar\phi-U(\bar\phi)\right]\,,
\end{equation}
where $U(\bar\phi)=\left(\mu^2\,\bar\phi^2/2+q\,\bar\phi/\tilde{c}\right)/\Omega^4$. The formulae simplify for $\xi=1/6$, in which case the canonically normalized scalar reads $\bar\phi_c=\sqrt{6}\,M_P\,{\mathrm {arctanh}}\,\left(\bar\phi/\sqrt{6}\,M_P\right)$, and has potential 
\begin{align}\label{nonminpot}
U(\bar\phi_c'\equiv\frac{\bar\phi_c}{\sqrt{6}M_P})=&\cosh^2\bar\phi_c'\,\sinh\bar\phi_c'\left(3\,\mu^2M_P^2\sinh\bar\phi_c'+\right.\nonumber\\
&+\left.\frac{q}{\tilde{c}}\,\sqrt{6}\,M_P\,\cosh\bar\phi_c'\right)\,.
\end{align}
The expression~\eqref{nonminpot} shows that, for non-minimal coupling to gravity, the dependence of the system on the integration constant $q$ becomes nontrivial. In this case, for instance, when $q$ is large enough the potential becomes unbounded from below.

\paragraph*{Discussion and conclusions.} We have shown that the Lagrangian~\eqref{laglong} is, despite its higher derivatives, a well-behaved Lagrangian: neither ghosts nor tachyons are propagating degrees of freedom. 

When we couple the form field to matter, instabilities do emerge, but, since they are non-perturbative, their characteristic time can be made safely long. Given the similarities between our system and the new massive gravity of~\cite{Bergshoeff:2009hq}, it would be interesting to inquire whether analogous instabilities appear in that model when gravity is coupled to matter. As we noted above, the absence of an absolutely stable minimum lies in the ``wrong'' sign of the would-be kinetic term for the $d$-form, {\it i.e.}, $F^{2}$; similarly in the model of~\cite{Bergshoeff:2009hq}, the would-be kinetic term for the gravitational degrees of freedom, {\it i.e.}, $R$, is chosen to have the ``wrong'' sign in order for the mass of the propagating degree of freedom to be non-tachyonic, hence we can speculate that a non-perturbative instability similar to the one we found for the present model could appear when the new massive gravity is coupled to matter.

As in our case,~\cite{Deffayet:2010zh} has also recently obtained a healthy behavior for form fields despite a higher derivative Lagrangian. Our mechanism is however different, as one can see by observing that~\cite{Deffayet:2010zh} considers form fields whose rank is different from the one we are considering.

Let us conclude with a speculation. The goal of this paper is to show an example of a higher derivative gauge theory that does not have any unhealthy behavior and that propagates massive degrees of freedom. Our goal is {\em not} to embed this theory in a UV-complete model; it would be interesting, however, to study whether the Lagrangian~\eqref{laglong} can emerge as a limit of some known theories, {\it e.g.}, as an effective low-energy description arising from strong dynamics at scale $M$. Within this framework, the model we presented might be a generalization of the models describing a condensation of topological defects, as studied in~\cite{Quevedo:1996uu}. 

{\bf Acknowledgments.} This work has been supported in part by the NSF grant PHY - 0855119. We thank John Donoghue, Nemanja Kaloper and Albion Lawrence for useful discussions.

\bibliography{agsForms10}

\end{document}